\documentclass{article}

\usepackage[utf8]{inputenc} 
\usepackage[T1]{fontenc}    
\usepackage{hyperref}       
\usepackage{url}            
\usepackage{booktabs}       
\usepackage{amsfonts}       
\usepackage{nicefrac}       
\usepackage{microtype}      
\usepackage{pgfplots}
\usepackage{tikz}
\usepackage{caption}

\usepgfplotslibrary{dateplot}
\pgfplotsset{compat=1.12}
   \newenvironment{customlegend}[1][]{%
        \begingroup
        \csname pgfplots@init@cleared@structures\endcsname
        \pgfplotsset{#1}%
    }{%
        \csname pgfplots@createlegend\endcsname
        \endgroup
    }%
    \def\addlegendimage{\csname pgfplots@addlegendimage\endcsname}

\newenvironment{myplot}[3][??]
{
\begin{tikzpicture}{center}
\begin{axis}[
scale only axis,
height=3cm,
width=\textwidth*#2,
grid=both,
max space between ticks=80,
minor x tick num=5,
minor y tick num=5,
major tick length=0.25cm,
minor tick length=0.1cm,
tick style={semithick,color=black},
xticklabel={\year-\month},
x tick label style={align=center}
]
\addplot+[no markers,thin] table [col sep=comma,trim cells=true,y=Groud_Truth#1] {#3};
\addplot+[no markers,thin] table [col sep=comma,trim cells=true,y=Forcast#1] {#3};
}
{
\end{axis}
\end{tikzpicture}
}

\title{\Large\bfseries Machine Learning Approaches for Traffic Volume Forecasting:
A Case Study of the Moroccan Highway Network}

\usepackage[T1]{fontenc}
\usepackage[utf8]{inputenc}
\usepackage{authblk}

\author[*]{Abderrahim KHALIFA \texttt{abderrahimkhalifa@student.emi.ac.ma}}
\author[*]{Younes IDSOUGUOU  \texttt{younesidsouguou@student.emi.ac.ma}}
\author[*]{Loubna BENABBOU \texttt{benabbou@emi.ac.ma} }
\author[**]{\break Mourad ZIRARI \texttt{zirari.mourad@adm.co.ma}}
\affil[*]{Department of Industrial Engineering, Ecole Mohammadia d'Ingenieurs \\
Mohammed V University, Rabat, Morocco}
\affil[**]{Department of Information Systems, Societe Nationale des Autoroutes du Maroc S.A\\
Rabat, Morocco
}

\begin{document}
\captionsetup{justification=centering}

\maketitle

\begin{abstract}
In this paper, we aim to illustrate different approaches we followed while developing a forecasting tool for highway traffic in Morocco. Two main approaches were adopted: Statistical Analysis as a step of data exploration and data wrangling. Therefore, a beta model is carried out for a better understanding of traffic behavior. Next, we moved to Machine Learning where we worked with a bunch of algorithms such as Random Forest, Artificial Neural Networks, Extra Trees, etc. yet, we were convinced that this field of study is still considered under state of the art models, so, we were also covering an application of Long Short-Term Memory Neural Networks.

\end{abstract}

\section{Introduction}

Highway traffic forecasting is a challenging research area in developing countries. The quality of forecasting is a key consideration as it leads to many useful applications, such as designing and upgrading highway networks, improving traffic safety, managing increasing congestion, and reducing alarming environmental pollution \cite{r1,r2,r3,r4,r5,r6}. Furthermore, the availability of accurate and reliable traffic information renders developing Intelligent Transporting Systems (ITS) more successful with a real impact for both transportation agencies and Highway users \cite{r1}. ITS can play a key role in developing country, by elaborating  applications related to the facilitation of transport to improve sustainable development outcomes \cite{r7}

Applying only classical statistical models, such as Autoregressive Integrated Moving Average (ARIMA), for forecasting by-passes important issues such as: (i) processing outliers, missing and noisy data  \cite{r8}. (ii) depicting high dimensional and non-linear relationship \cite{r9} iii) spatiotemporal evolution of traffic \cite{r10,r11,r12,r13}. In this paper we take a complementary stance and we apply machine learning algorithms for predicting traffic in Moroccan Highway to illustrate the benefits of applying Machine Learning techniques for forecasting in developing country. Section 2 presents the case study with data exploration for a better understanding of the nature of data and the traffic behavior. Section 3 focuses on the presentation of used machine learning methods. Application and results interpretation will be discussed in section 4.

\section{Problem description}

Between 2008 and 2015, the Moroccan highway network has known a great growth going from 850 km to 1400 km, and it will be more likely to grow in the few coming years. So as to do up with the increasing transport demand, knowing we have a 4 years of daily historical traffic volumes data at each network's toll station, we aim to predict on what would be the daily traffic volumes for the next two years. We have 3 classes of vehicles:

\begin{itemize}

\item \textbf{TC1: }Light and small vehicles (Cars, Bikes)
\item \textbf{TC2: }Medium vehicles (Vans, Trucks ...)
\item \textbf{TC3: }Heavy vehicles (Trailer trucks, Buses ...)

\end{itemize}

In the data exploration, we noticed the presence of some anomalies on the data. There were some abnormal peaks and trends and in some cases the absence of traffic for a period of time. We explain these strange behaviors in data with many factors. First of it, by matching the data with the calendar we notice some national events where the traffic is influenced right before and after the event which results in a instant trend that collapses gradually. Also, we found some other punctual phenomenon that affects the traffic on a specific date which results in peaks or absence of traffic for a given period, i.e New Year's Eve.

In order to deal with these anomalies, we adopted a data filter after trying several methods like : Moving average, Exponential moving average, and Deseasonalization.

\begin{figure}[h]
\begin{tikzpicture}{center}
\begin{axis}[
scale only axis,
height=4cm,
width=\textwidth*0.9,
grid=both,
max space between ticks=80,
minor x tick num=5,
minor y tick num=5,
major tick length=0.25cm,
minor tick length=0.1cm,
tick style={semithick,color=black},
date coordinates in=x,
xticklabel={\year},
x tick label style={align=center},
date ZERO=2012-01-01,
xmin={2013-01-01},
xmax={2017-01-01},
ymin={0}
]
\addplot+[no markers,thin] table [col sep=comma,trim cells=true,y=value] {rawdata.txt};
\addplot+[no markers,thick] table [col sep=comma,trim cells=true,y=value] {processeddata.txt};
\addlegendentry{Raw daily traffic}
\addlegendentry{Daily traffic data with median filter}

\end{axis}
\end{tikzpicture}
  \caption{daily traffic without smoothing and its correction with the median filter}
\end{figure}
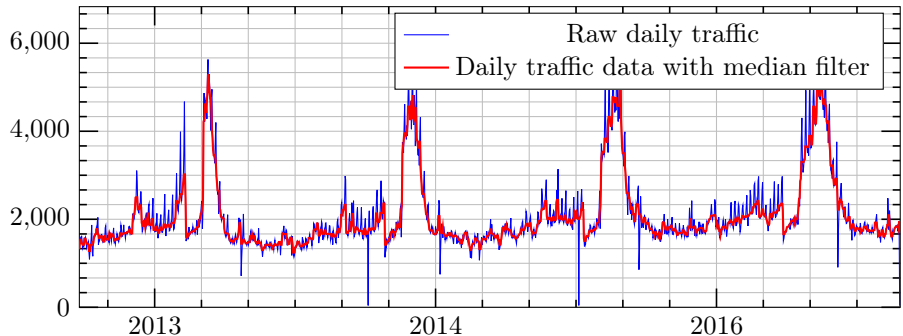

This filter is median based on a window of 5 days, and from the figure above we see its robustness against outliers.

\section{Used methods}
Our first attempt was mainly based on statistical models (\verb+Holt-Winters+, \verb+ARIMA+ ...), but it comes to give a poor performance on most stations. This was due to the high nonseasonal variations in the data that these traditional models couldn't explain.

Next, we set out to move on something new and more advanced in modeling. We started out by adopting some machine learning techniques after trying several algorithms and figured out the ones with the best accuracy rates. We short-list here the algorithms used in this study.

\textbf{Random Forest: }It is a supervised classification algorithm. As the name suggest, this algorithm creates the forest with a number of trees. In general, the more trees in the forest the more robust the forest looks like. In the same way in the random forest classifier, the higher the number of trees in the forest gives the high accuracy results. Its main parameters are : the number of trees in the forest, the number of features to consider when looking for the best split, the minimum number of samples needed to split an internal node and the minimum number of samples required to be at a leaf node \cite{r14}.

\textbf{Extra trees: }Also known as the extremely randomized trees it's a tree-based method for supervised classification where splits are performed totally or partially at random \cite{r15}.

\textbf{Multi-layer Perceptron: }An MLP is a network of simple neurons called perceptrons. The perceptron computes a single output from multiple real-valued inputs by forming a linear combination according to its input weights and then possibly putting the output through some nonlinear activation function. Multi-layer Perceptron is sensitive to feature scaling, so it is highly recommended to scale the data, like scaling each attribute on the input vector $X \in [0, 1]$ or $[-1, +1]$, or standardize it to standard normal distribution. Its parameters are : The regularization parameter, The number of hidden layers and number of neurons within each layer and Maximum number of iterations \cite{r17,r18}.

\textbf{AdaBoost: }It can be used in conjunction with many other types of learning algorithms to improve their performance. The output of the other learning algorithms \verb+weak learners+ is combined into a weighted sum that represents the final output of the boosted classifier. its parameters are : The number of estimators and learning rate \cite{r19,r20}.

Before starting to use these algorithms we manipulated the data so as to design
the learning matrix which will be the input of the used algorithms. To remind
the initial data we had : \verb+YEAR+, \verb+MONTH+, \verb+DATE+, \verb+TC1+, \verb+TC2+, \verb+TC3+.

So as to make this matrix much more interesting to a learning algorithm we added
some extra columns by exploding the column \verb+date+ as shown in the following table:

\begin{table}[h]
  \caption{Data Structure}
  \resizebox{\textwidth}{!}{%
  \label{data-structure-table}
  \centering
  \begin{tabular}{llllllll}
    \toprule
     \multicolumn{8}{c}{Attributes} \\
    \midrule
     STATION\_CODE &YEAR & MONTH & WEEK & DAY & DAY\_OF\_WEEK & DAY\_OF\_YEAR & IS\_WEEKEND\\
     Integer & Integer & Integer & Integer & Integer & Integer & Integer & Boolean\\
    \bottomrule
  \end{tabular}}
\end{table}

Even if the \verb+DAY_OF_WEEK+ attribute seems to be integer but it's somehow a categorical variable that is indicating the day of the week using a number ranging from 0 to 6. Learning with a column like this one, supposes that the days influence the traffic in an equal way which isn't right. Then, in order to let the algorithm to learn the impact of each day on the outcome, we consider some additional dummy variables (7 variables) which correspond to a given day of the week.

\textbf{Long Short Term Memory networks} or \verb+LSTM+ are a special kind of Reccurent Neural Networks, capable of learning long-term dependencies. LSTMs are explicitly designed to avoid the long-term dependency problem. Remembering information for long periods of time is practically their default behavior, not something they struggle to learn \cite{r21}.

The previous approach described above is not compatible with LSTM Networks. Therefore, we have to reconstruct our learning matrix. All the columns we used before seems to be not useful at this stage since with LSTM we only need the historical traffic values on a given station. We have two kinds of forecasts using LSTM: One-step forecasts where we predict the next value for a serie of measures, and Multi-steps forecasts where we predict multiple values at a time. Both techniques requires as an input the past measured values within a lookback range.

At this stage, we explained that the statistical models were unable to give good prediction results. That's why we moved to machine learning techniques that are somehow more sophisticated. These algorithms need a well designed learning matrix. It was done by exploding the date attribute to many explanatory variables and also by adding some dummy variables at categoricals in order to give the model more flexibily. Finally, we tried out the LSTM neural network that is also good at predicting time series especially with the underlying memory kept behind. It requires also a specific learning matrix design which is focused on a shifted target variable measure at the input. Next, we will all apply these algorithm and find out which one is give the best results.

\section{Application}
In this section, we will be using algorithms discussed in the previous part to forecast the volumes. We focused only on the "light vehicles" class to make
predictions on a given toll station. So we will be presenting resulting prediction graph for each algorithm with it's actual parameters.

\begin{figure}[h]
\centering
\begin{tikzpicture}{center}
\begin{axis}[
scale only axis,
height=2.5cm,
width=\textwidth*0.8,
grid=both,
max space between ticks=80,
minor x tick num=5,
minor y tick num=5,
major tick length=0.25cm,
minor tick length=0.1cm,
tick style={semithick,color=black},
date coordinates in=x,
xticklabel={\year-\month},
x tick label style={align=center},
date ZERO=2016-06-01,
xmin={2016-06-01},
xmax={2017-05-01},
ymin={0}
]
\addplot+[no markers,thin] table [col sep=comma,trim cells=true,y=TC1] {daily_ANN1.csv};
\addplot+[no markers,thin] table [col sep=comma,trim cells=true,y=TC1Predit] {daily_ANN1.csv};

\addlegendentry{Measured traffic}
\addlegendentry{ANN forecast}

\end{axis}
\end{tikzpicture}
  \caption{Traffic forecasting using MLP (hidden\_layer\_sizes=(200,100,100,200,100,200), alpha=0.01, max\_iter=80000, batch\_size=40)}
\end{figure}
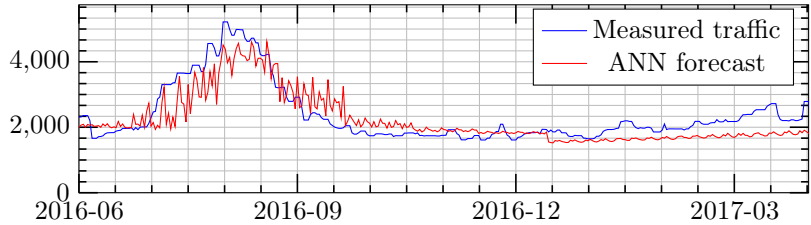

\begin{figure}[h]
\centering
\begin{tikzpicture}{center}
\begin{axis}[
scale only axis,
height=2.5cm,
width=\textwidth*0.8,
grid=both,
max space between ticks=80,
minor x tick num=5,
minor y tick num=5,
major tick length=0.25cm,
minor tick length=0.1cm,
tick style={semithick,color=black},
date coordinates in=x,
xticklabel={\year-\month},
x tick label style={align=center},
date ZERO=2016-06-01,
xmin={2016-06-01},
xmax={2017-05-01},
ymin={0}
]
\addplot+[no markers,thin] table [col sep=comma,trim cells=true,y=TC1] {dailyExtraTrees.csv};
\addplot+[no markers,thin] table [col sep=comma,trim cells=true,y=TC1Predit] {dailyExtraTrees.csv};

\addlegendentry{Measured traffic}
\addlegendentry{Extra trees forecast}

\end{axis}
\end{tikzpicture}
  \caption{Traffic forecasting using ExtraTrees (n\_estimators=100, min\_sample\_split=10, min\_sample\_leaf=13, max\_features=0.75*n\_features)}
\end{figure}
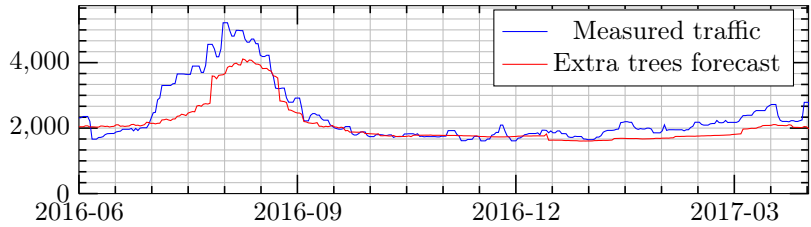

\begin{figure}[h]
\centering
\begin{tikzpicture}{center}
\begin{axis}[
scale only axis,
height=2.5cm,
width=\textwidth*0.8,
grid=both,
max space between ticks=80,
minor x tick num=5,
minor y tick num=5,
major tick length=0.25cm,
minor tick length=0.1cm,
tick style={semithick,color=black},
date coordinates in=x,
xticklabel={\year-\month},
x tick label style={align=center},
date ZERO=2016-06-01,
xmin={2016-06-01},
xmax={2017-05-01},
ymin={0}
]
\addplot+[no markers,thin] table [col sep=comma,trim cells=true,y=TC1] {daily_RandomForest.csv};
\addplot+[no markers,thin] table [col sep=comma,trim cells=true,y=TC1Predit] {daily_RandomForest.csv};

\addlegendentry{Measured traffic}
\addlegendentry{RandomForest forecast}

\end{axis}
\end{tikzpicture}
  \caption{Traffic forecasting using RandomForest (n\_estimators=5000, min\_sample\_split=2, min\_sample\_leaf=1, max\_features=sqrt(n\_features))}
\end{figure}
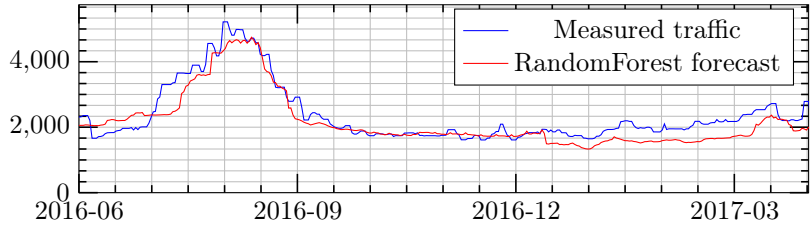

Plots for Adaboost and LSTM application are listed on the Appendix.

\break
Now that we have the predictions using various algorithms, it's time to compare the performance of each algorithm and choose the best one. We have here a table comparing accuracy of each model.

\begin{table}[h]
  \caption{Performace summary based on SMAPE metric}

  \label{data-structure-table}
  \centering
  \begin{tabular}{lllll}
    \toprule
     \multicolumn{5}{c}{Models} \\
    \midrule
     MLP &RandomForest & AdaBoost & ExtraTrees & LSTM\\
     17\% & 13.4\% & 15\% & 12.8\% & 19\% \\
    \bottomrule
  \end{tabular}
\end{table}

We can figure out now that all these models are somehow interesting. So as to increase their performance, we need to get through some advanced techniques. That is to
say making combination of them using "Ensemble Methods" (Bagging, Boosting, Stacking).

\section*{Appendix}

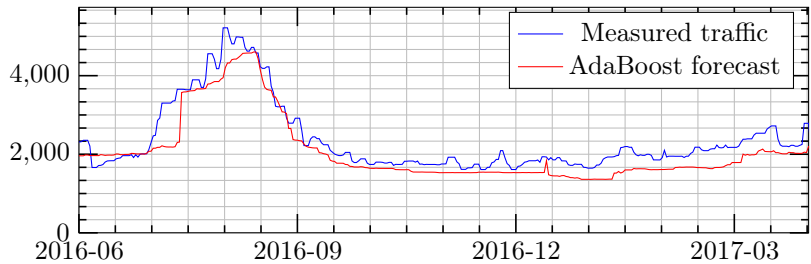
\begin{figure}[!h]
\centering
\begin{tikzpicture}{center}
\begin{axis}[
scale only axis,
height=3cm,
width=\textwidth*0.8,
grid=both,
max space between ticks=80,
minor x tick num=5,
minor y tick num=5,
major tick length=0.25cm,
minor tick length=0.1cm,
tick style={semithick,color=black},
date coordinates in=x,
xticklabel={\year-\month},
x tick label style={align=center},
date ZERO=2016-06-01,
xmin={2016-06-01},
xmax={2017-05-01},
ymin={0}
]
\addplot+[no markers,thin] table [col sep=comma,trim cells=true,y=TC1] {dailyAdaBoost.csv};
\addplot+[no markers,thin] table [col sep=comma,trim cells=true,y=TC1Predit] {dailyAdaBoost.csv};

\addlegendentry{Measured traffic}
\addlegendentry{AdaBoost forecast}

\end{axis}
\end{tikzpicture}
  \caption{Traffic forecasting using AdaBoost (n\_estimators=10000, learning\_rate=1)}
\end{figure}

\sbox0{\begin{myplot}[8]{0.4}{lstm.csv}
\end{myplot}}

\sbox1{\begin{myplot}[4]{0.4}{lstm.csv}
\end{myplot}}

\sbox2{\begin{myplot}[13]{0.4}{lstm.csv}
\end{myplot}}

\sbox3{\begin{myplot}[7]{0.4}{lstm.csv}
\end{myplot}}

\begin{minipage}{\dimexpr \wd0+\wd1}
\centering
\usebox0\usebox1
\usebox2\usebox3
\begin{tikzpicture}
\begin{customlegend}[legend columns=5,legend style={align=center,draw=none,column sep=2ex},
        legend entries={Measured traffic,
                        Forcasted traffic ,
                        }]
        \addlegendimage{mark=none,solid,thin,line legend,blue}
        \addlegendimage{mark=none,solid,thin,red}
        \end{customlegend}
\end{tikzpicture}

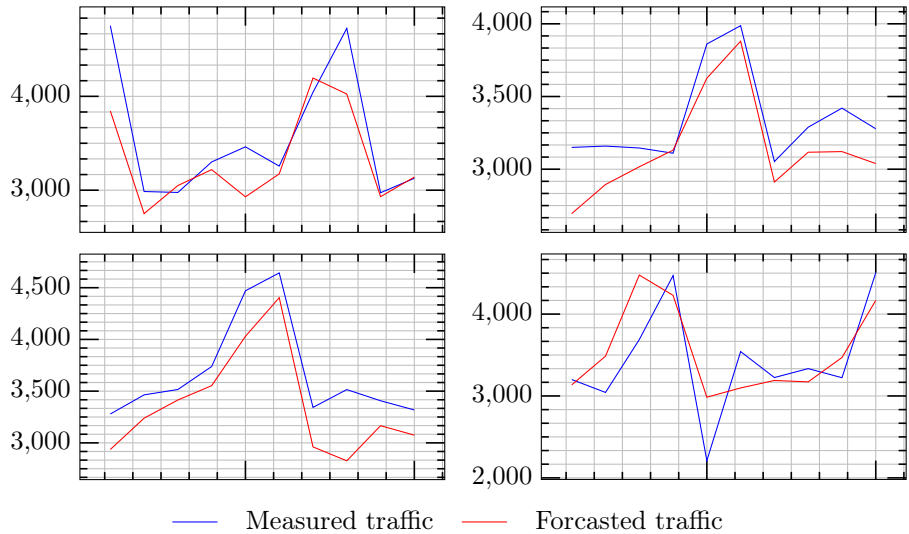
\captionof{figure}{Traffic forecasting using LSTM with 10 steps ahead, LSTM (Epoch=1000, LSTM\_hidden\_layer\_sizes=(100,100,100,100), activation="linear", lookback=100, lookforward=10)}
\end{minipage}

\end{document}